\documentclass[reprint,showpacs,aps,pra]{revtex4-1}
\usepackage{graphics}
\usepackage{bm}
\usepackage{amsmath}
\usepackage{amssymb}
\usepackage{graphicx}
\usepackage{amsthm}
\begin{document}

\title{One-body information loss in fermion systems}
\author{N. Gigena, R.Rossignoli} 
\affiliation{IFLP-Departamento de F\'{\i}sica, 
Universidad Nacional de La Plata, C.C. 67, La Plata (1900), Argentina}

\begin{abstract}
We propose an  entropic measure of non-classical correlations in general mixed
states of fermion systems, based on the loss of information due to the unread
measurement of the occupancy of single particle states of a given basis. When
minimized over all possible single particle bases, the measure reduces to an
entanglement entropy for pure states and vanishes only for states which are
diagonal in a Slater determinant basis. The approach is also suitable for
states having definite number parity yet not necessarily a fixed particle
number, in which case the minimization can be extended to all bases related
through a Bogoliubov transformation if quasiparticle mode measurements are also
considered. General stationary conditions for determining the optimizing basis
are derived. For a mixture of a general pure state with the maximally mixed
state, a general analytic evaluation of the present measure and optimizing
basis is provided, which shows that non-entangled mixed states may nonetheless
exhibit a non-zero information loss.
\pacs{03.67.Mn, 03.65.Ud, 05.30.Fk}

\end{abstract} 
\maketitle

\section{Introduction}

Quantification of non-classical correlations in quantum systems is one of the
main topics in quantum information theory. Quantum entanglement is the most
famous and best studied manifestation of such correlations, mainly because of
its central role as a resource for quantum teleportation \cite{Be.93} and
quantum computation \cite{NC.00, JLV.03,VV.03}. Nevertheless, quantum
entanglement is not the only type of non-classical correlation.  It is now
well-known that non-entangled mixed states can also exhibit non-classical
features, which may be relevant  in mixed-state based algorithms such as that
of Knill and Laflamme \cite{KL.98,DFC.05}. A great effort has therefore been
devoted in recent years to understand and quantify the quantumness of
correlations \cite{KM.12}. Various measures of these correlations beyond
entanglement have then been proposed, which start with the quantum discord
\cite{OZ.01,HV.01,Z.03,DSC.08} and the one-way information deficit
\cite{OH.02,SKB.11}. The latter has been extended to more general
entropic forms \cite{RCC.10}, as a measure of the information loss due to a
local measurement. Various other related measures have also been later
introduced \cite{KM.12,DKV.10,PGA.11,GTA.13,LFO.12,POS.13,NPA.13,CTG.14}.

While these correlation measures have been intensively investigated in systems
of distinguishable components, less attention has been given to their
extension to systems of identical particles \cite{IDV.14}, and in particular to
fermion systems.  In these systems particles cannot be accessed individually
because of indistinguishability, thus preventing the straightforward extension
of the correlation measures defined for systems of distinguishable components,
which are based on the tensor product structure of the state space. This property
no longer holds for indistinguishable components, turning the characterization of
correlations more complex. Various approaches for describing and quantifying
entanglement in fermion systems have been introduced, based on
quantum correlations and particle-like entanglement
\cite{Sch.01,SDM.01,Eck.02,Wi.03,Ci.09,IV.13,Ci.09,Os.14,SL.14,GR.15} and also
on mode-type entanglement \cite{Za.02,Shi.03,Fri.13,Be.14}.

In a previous work \cite{GR.15} we analyzed the problem of quantifying
entanglement in pure and mixed states of fermion systems having definite number
parity, yet not necessarily a fixed particle number. The approach is based on a
consistent definition of the measurement of the occupancy of a single-particle
(sp) mode, and of the reduced state of such a mode and its orthogonal
complement in the sp Hilbert space $\mathcal{S}$. It leads to an entanglement
entropy which is explicitly invariant under particle-hole (and also Bogoliubov)
transformations, and which extends previous treatments for states with fixed
particle number \cite{Sch.01}, providing at the same time a link between
mode-based and particle based approaches.

In this work we start by considering again the correlations between a sp mode
and its orthogonal complement and define an entropic measure that quantifies
the loss of information in the state of the system due to the measurement of
the occupancy of that mode. Such loss is directly related to the entanglement
generated between the system and the measurement device. We then consider the
sum over all states on a given basis of $\mathcal{S}$, minimized over all such
bases, of this quantity as a measure of the minimum loss of information due to
an unread measurement in this basis, extending the minimization to all bases
related through a Bogoliubov transformation \cite{RS.80} if the occupancy
measurement of quasiparticle modes is also allowed. It is then shown that this
measure is a non negative quantity which is zero if and only if the state of
the system is a convex combination of Slater determinants in the same sp basis,
reducing for pure states to the entanglement entropy defined in \cite{GR.15},
in analogy with its counterpart \cite{RCC.10} for systems of distinguishable
constituents. General stationary conditions for the optimizing basis are also
derived. An analytic evaluation of this measure for a  mixture of a general
pure state with the maximally mixed state is provided, which shows that in this
case the optimizing quasiparticles are just those  diagonalizing the
generalized one-body density matrix, and that the present measure can be
non-zero in non-entangled mixed states. Explicit comparison with the fermionic
entanglement of formation in a specific case is also made. Two fermion states
are as well discussed.

\section{Formalism}

\subsection{Single mode measurement and entanglement entropy}

We first consider a pure state $|\psi\rangle$ of a fermion system with an
$n$-dimensional single-particle (sp) Hilbert space $\mathcal{S}$. The system is
described by a set of operators $\{c_j, c^\dagger_j,\}$ satisfying the
canonical anti-commutation relations
\begin{equation}
\{c_i,c_j\}=0,\;\;\{c_i,c_j^\dagger\}=\delta_{ij},
\end{equation}
such that $\{|j\rangle=c^\dagger_j|0\rangle,\;j=1,\ldots,n\}$ is an orthonormal set of
one fermion states states ($|0\rangle$ denotes the vacuum of the operators
$c_j$). We will work within a grand canonical context, so the state
$|\psi\rangle$ does not necessarily has a definite particle number $N=\sum_k
c^\dagger_kc_k$. Nonetheless, we will always assume that  $|\psi\rangle$ has a
definite number parity,
\begin{equation}
P|\psi\rangle=\pm|\psi\rangle,\;\;P=\exp[i\pi N]= \prod_k(1-2c^\dagger_k c_k)\,,\label{P}
\end{equation}
in agreement with the standard superselection rule \cite{Fr.16}.
We will also denote by $d$ the dimension of the Fock space of the system.

We now consider a partition $(A,B)$ of a particular basis of $\mathcal{S}$ (see Appendix),
where $A$ denotes the single mode or ``level'' $k$ and $B$ its orthogonal
complement. Due to the anticommutation relations, the operators
\begin{equation}
\Pi_k=c^\dagger_k c_k,\;\;\Pi_{\bar{k}}=c_k c^\dagger_k\,,\;\;\Pi_k
+\Pi_{\bar{k}}=1\label{4},
\end{equation}
constitute a basic set of orthogonal projectors, defining a standard projective
measurement on the level $k$. The operator $\Pi_k$ ($\Pi_{\bar k}$) projects
the state onto the subspace of states with level $k$ occupied (empty), so the
set describes the measurement of the occupancy of this level. The ensuing post
measurement states are $|\psi_k\rangle=\Pi_k|\psi\rangle/\sqrt{p_k}$ and
$|\psi_{\bar k}\rangle=\Pi_{\bar k}|\psi\rangle/\sqrt{p_{\bar k}}$, with
$p_k=\langle c^\dagger_k c_k\rangle=\langle \psi|c^\dagger_k c_k|\psi\rangle$
and $p_{\bar k}=\langle c_k c^\dagger_k\rangle=1-p_k$, such that
\begin{equation}|\psi\rangle=\sqrt{p_k}|\psi_k\rangle+\sqrt{p_{\bar{k}}}|\psi_{\bar{k}}\rangle
\label{psik}\,.\end{equation}
For any operator $O_A$ ($O_B$) that depends only on the operators $c_k,
c^\dagger_{k}$ ($\{c_j,c^\dagger_j,j\neq k\}$) we have, using (\ref{P}),
\begin{eqnarray}
\langle\psi| O_{A(B)}|\psi\rangle&=& p_k \langle\psi_k|O_{A(B)}|\psi_k\rangle
+p_{\bar k}\langle\psi_{\bar k}|O_{A(B)}|\psi_{\bar k}\rangle\nonumber\\
 &=&{\rm Tr}\,\rho_{A(B)}O_{A(B)} \,,\end{eqnarray}
where
\begin{eqnarray}\rho_A&=&p_k  c^\dagger_k|0\rangle\langle 0|c_k +p_{\bar k} |0\rangle\langle 0|
\label{rhoA}\,,\\
\rho_B&=&p_k  c_k|\psi_k\rangle\langle \psi_k|c^\dagger_k +p_{\bar k}
|\psi_{\bar k}\rangle\langle \psi_{\bar k}|\,.\label{rhoB}\end{eqnarray}
 play the role of reduced states for $A$ and $B$.

As shown in \cite{GR.15}, for pure states the mode measurement defined above
allows to define an entanglement entropy that quantifies the entanglement
between the measured sp mode and its orthogonal complement in $\mathcal{S}$.
Indeed, the reduced states (\ref{rhoA})--(\ref{rhoB}) have the same
eigenvalues $\{p_k,p_{\bar k}\}$, and hence the entanglement entropy for
such bipartition is the Shannon entropy of that distribution:
\begin{equation}
E_k(|\psi\rangle)=-p_k\log_2(p_k)-p_{\bar k}\log_2(p_{\bar k})\label{Ek}\,.
\end{equation}
The sum of (\ref{Ek}) over all states on a given basis of $\mathcal{S}$,
minimized over all sp bases, is the \emph{one-body entanglement entropy}
\cite{GR.15}
\begin{equation}
S^{\rm sp}(|\psi\rangle)=\mathop{\rm Min}_{\{c_k\}}
\sum_k -p_k\log_2 p_k -p_{\bar k}\log_2 p_{\bar k}\label{ent}\,.
\end{equation}
The minimum in Eq.\ (\ref{ent}) is reached for those operators $\{c_k\}$ that
diagonalize the one-body density matrix $\rho_{ij}^{\rm sp}=\langle
{c}_j^\dagger {c}_i\rangle$, i.e., for those satisfying $\langle {c}_j^\dagger
{c}_i\rangle=\lambda_i\delta_{ij}$ \cite{GR.15}. Hence,
\begin{equation}
S^{\rm sp}(|\psi\rangle)={\rm tr}\,h(\rho^{\rm sp}),
\end{equation}
where $h(p)=-p\log_2 p-(1-p)\log_2(1-p)$ and  tr the trace in the sp space.  It is clear then that $S^{\rm sp}(|\psi\rangle)=0$ iff there is a sp basis in which the state is written as a
Slater determinant, for in that case the eigenvalues of $\rho^{\rm sp}$ are 
either 0 or 1 ($(\rho^{\rm sp})^2=\rho^{\rm sp}$). Then Slater determinants are
considered here non-correlated states, in agreement with \cite{Sch.01}.

If quasiparticle modes are to be allowed, then the minimization extends to all
quasiparticle bases, i.e., sets of operators $\{a_i, a^\dagger_i\}$ related to
the original fermion operators $\{c_i, c^\dagger_i\}$ through a Bogoliubov
transformation \cite{RS.80}:
\begin{equation}
 a_i=\sum_j \bar{U}_{j i} c_j+V_{j i}c^\dagger_j\,. \label{anu}
\end{equation}
Eq.\ (\ref{anu}) can be written as
\begin{equation}
\left(\begin{array}{c}\bm{a}\\\bm{a}^\dagger\end{array}\right)={\cal W}^\dagger
\left(\begin{array}{c}\bm{c}\\\bm{c}^\dagger\end{array}\right)\,, \;\;{\cal
W}=\left(\begin{array}{cc}U&V\\\bar{V}&\bar{U}\end{array}\right)\,,\label{Bg}
\end{equation}
where the $2n\times 2n$ matrix ${\cal W}$ should be unitary
($UU^\dagger+VV^\dagger=1,\;\;UV^T+VU^T=0$) in order that $a_\nu,a^\dagger_\nu$
fulfill the fermionic anticommutation relations (1). It can be  shown
\cite{GR.15} that the minimum is reached for those operators $\{a_\mu,
a^\dagger_\mu\}$ that diagonalize the {\it extended} $2n\times 2n$ density
matrix
\begin{equation}
\rho^{\rm qsp}= 1-\langle\left(\begin{array}{c}\bm{c}\\
\bm{c}^\dagger\end{array}\right)\left(\begin{array}{cc}\
\bm{c}^\dagger&\bm{c}\end{array}\right)\rangle=
\left(\begin{array}{cc}\rho^{\rm sp}&\kappa\\-\bar{\kappa}\;\;&
1-\bar{\rho}^{\rm sp}\end{array}\right)\label{qsp}\,,
\end{equation}
where $\kappa_{ij}=\langle c_jc_i\rangle$,  $-\bar{\kappa}_{ij}=\langle
c^\dagger_j c^\dagger_i\rangle$ and $(1-\bar{\rho}^{\rm sp})_{ij}=\langle c_j
c^\dagger_i\rangle$. Eq.\ (\ref{qsp}) is an hermitic matrix which can always be
diagonalized by a suitable transformation (\ref{Bg}), such that
\[
1-\langle\left(\begin{array}{c}\bm{a}\\
\bm{a}^\dagger\end{array}\right)\left(\begin{array}{cc}\bm{a}&\bm{a}^\dagger\end{array}\right)\rangle
={\cal W}^\dagger\rho^{\rm qsp}{\cal W}= \left(\begin{array}{cc}f&0\\ 0&1-f\end{array}\right)\,,
\]
with $f_{kl}=f_k\delta_{kl}$ and $f_k$, $1-f_k$  the eigenvalues
of $\rho^{\rm qsp}$ (which come in pairs $(f_k,1-f_k)$, with
$f_k\in[0,1]$), entailing $\langle a^\dagger_k a_k\rangle=\delta_{kl}
f_k$, $\langle a_k a_l\rangle=0\,.$ The generalized one-body
entanglement entropy is therefore
\begin{eqnarray}
S^{\rm qsp}(|\psi\rangle)&=&-\sum_k f_k\log_2 f_k+(1-f_k)\log_2(1-f_k)\label{666}\notag\\
&=&-{\rm tr}'\,\rho^{\rm qsp}\log_2 \rho^{\rm qsp}\,,\label{sqsp2}
\end{eqnarray}
with ${\rm tr}'$ the trace in the extended single particle space. Hence,
$S^{\rm qsp}(|\psi\rangle)=0$ iff all eigenvalues $f_k$ are $0$ or $1$, i.e.,
$(\rho^{\rm qsp})^2=\rho^{\rm qsp}$, which implies that $\rho$ is a vacuum or
in general a quasiparticle Slater Determinant.

\subsection{Information loss due to a single mode measurement}
Let us now consider a single mode measurement on a general mixed state $\rho$
of a fermion system (assumed to satisfy $[\rho,P]=0$). The state after finding
that sp state $|k\rangle$ is occupied or empty is
$\rho'_k=\Pi_k\,\rho\,\Pi_k/p_k$, or $\rho'_{\bar k}=\Pi_{\bar
k}\,\rho\,\Pi_{\bar k}/p_{\bar k}$ respectively, with $p_k=\langle c^\dagger_k
c_k\rangle={\rm Tr}\,\rho\,c^\dagger_k c_k$ and $p_{\bar{k}}=1-p_k$. Therefore,
after an unread measurement of that level the state of the system is
\begin{eqnarray}
\rho'(k)=p_k \,\rho'_k+p_{\bar k}\,\rho'_{\bar k}=\Pi_{k}\,\rho\,\Pi_{k}
+\Pi_{\bar k}\,\rho\,\Pi_{\bar k}.\label{rhopk}
\end{eqnarray}
For instance, if $\rho$ is a pure state $|\psi\rangle\langle\psi|$ (Eq.\ (\ref{psik})),
\begin{equation}
\rho'(k)=p_k|\psi_k\rangle\langle\psi_k|+p_{\bar{k}}|\psi_{\bar{k}}\rangle\langle\psi_{\bar{k}}|\,.
 \label{rhoppk}\end{equation}
The ensuing reduced state of $A$ is given by (\ref{rhoA}) while $\rho_B= p_k\,
c_k\rho'_k c^\dagger_k + p_{\bar k}\, \rho'_{\bar k}$. Since
$\rho=\rho'(k)+\Pi_k\rho\Pi_{\bar{k}}+\Pi_{\bar{k}}\rho\Pi_k$, the last two
terms are lost after such measurement.

The entropy $S(\rho'(k))=-{\rm Tr}\,\rho'(k)\log_2\rho'(k)$ of the
post-measurement state (\ref{rhopk}) then cannot be lower than the entropy $S(\rho)$
of the original state, due to the information contained in the lost elements.
In fact, the eigenvalues of $\rho'(k)$ are just the diagonal elements of
$\rho$ in a basis different from that of its eigenvectors, and it is well known
that any such diagonal is always majorized by the eigenvalues of $\rho$
\cite{Bh.97}. We have then
\begin{equation}
S(\rho'(k))\geq S(\rho),
\end{equation}
with equality if and only if $\rho'(k)=\rho$. This last condition is obviously
equivalent to
\begin{equation}
[\rho,c^\dagger_k c_k]=0\,,\label{ck}
\end{equation}
since if $\rho=\rho'(k)$,  Eq.\ (\ref{ck}) holds, while if (\ref{ck}) is valid,
there is common basis of eigenvectors of $\rho$ and $c^\dagger_k c_k$ and hence
$\rho=\rho'(k)$. The difference
\begin{equation}
I^{c_k}(\rho)=S(\rho'(k))-S(\rho)\label{sld}
\end{equation}
quantifies then this loss of information. It clearly satisfies
$I^{c_k}(\rho)\geq 0$,  with  $I^{c_k}(\rho)= 0$ iff (\ref{ck}) applies. It is
a fermionic version of the information deficit
$\Delta^\rightarrow(\rho)=S(\rho')-S(\rho)$ \cite{OH.02}, defined for systems
of distinguishable constituents (there is here no minimization involved because
the occupation of mode $|k\rangle$ is a classical variable). Furthermore, it is
worth noting from (\ref{rhopk})  that
\begin{equation}S(\rho'(k))=S(k)+S(\rho|k)\label{spk},\end{equation}
where
\begin{equation}S(k)=-p_k\log_2 p_k-p_{\bar k}\log_2 p_{\bar k}=h(p_k)\end{equation}
is the entropy of mode $k$ and
\begin{equation}S(\rho|k)=p_k S(\rho'_k)+p_{\bar k} S(\rho'_{\bar k})\,,
\label{Sck}\end{equation}
is the conditional entropy of the set of remaining modes, given that the state
of mode $k$ (occupied or empty) is known. Therefore, the deficit (\ref{sld})
can be written as
\begin{equation}
S(\rho'(k))-S(\rho)=S(\rho|k)-(S(\rho)-S(k)),
\end{equation}
which is a difference of classical-like and quantum conditional entropies with
respect to mode $k$ and represents the quantum discord \cite{OZ.01} with
respect to this mode. Hence, for single-mode measurements the extension of
quantum deficit coincides with that of the quantum discord. For a pure state
$\rho=|\psi\rangle\langle\psi|$, $\rho'_k$ and $\rho'_{\bar k}$ are both pure
(Eq.\ (\ref{rhoppk})), implying that $I^{c_k}$ becomes coincident  with the
entanglement entropy (\ref{Ek}) of mode $k$:
\begin{equation}
I^{c_k}(|\psi\rangle)=S(k)=E_k(|\psi\rangle)\,.\label{Ipk}
\end{equation}

The conditional entropy (\ref{Sck}) can be interpreted, following the general
results of \cite{WK.04}, as an entanglement of formation between the set of
remaining modes $k'\neq k$ and a complementary system  (which can also be a set
of new fermionic modes) which purifies the whole system (see Appendix, Eq.\
(\ref{Eb1})). Besides, the arguments of \cite{SKB.11,PGA.11} imply that the
information deficit (\ref{sld}) is an indicator  of the entanglement generated
between the measurement device and the system after a measurement of mode $k$.
In fact, by adding a qubit ancilla $C$ in an initial state $|0\rangle$ to the
fermionic system and performing the unitary transformation
$U=e^{-i\frac{\pi}{2} (c_kc^\dagger_k)\otimes \sigma_y}$, we obtain
\begin{equation}
 U(|\psi\rangle|0\rangle)=
\sqrt{p_k}|\psi_k\rangle|0\rangle+\sqrt{p_{\bar k}}|\psi_{\bar k}\rangle|1\rangle\,.
\end{equation}
Hence, for a general mixed fermion state $\rho$,
\begin{equation}
 {\rm Tr}_C[U(\rho\otimes |0\rangle\langle 0|)U^\dagger]=\rho'(k)\,.
 \end{equation}
Therefore $I^{c_k}(\rho)$ is the difference between the entropy of the
fermionic subsystem $\rho_F={\rm Tr}_C\,\rho_{FC}=\rho'(k)$ in
$\rho_{FC}=U(\rho\otimes |0\rangle\langle 0|)U^\dagger$ and that of the whole
system, $S(\rho_{FC})=S(\rho)$. Such difference (the negative of the
conditional entropy $S(\rho_{FC})-S(\rho_F)$) can be positive only if
$\rho_{FC}$ is entangled, according to the entropic separability criterion
\cite{HH.96,NK.01,RC.02}, and is a lower bound to the one-way distillable
entanglement \cite{SKB.11,PGA.11,PV.07}. On the other hand, if 
$\rho'(k)=\rho$, such that $p_k=0$ or $1$ for any eigenstate of $\rho$, then 
$\rho_{FC}$ is clearly separable and no entanglement is created. 

\subsection{One-body information loss}
We now take the sum, over all the states of a given basis of $\mathcal{S}$, of
$I^{c_k}$ in (\ref{sld}),
\begin{equation}
I^{\bm{c}}(\rho)=\sum_k I^{c_k}(\rho)=\sum_kS(\rho'(k))-S (\rho)\,,
\end{equation}
as a measure of the loss of information due to an unread measurement in this
basis. The minimum over all sp bases of this difference,
\begin{equation}
I^{\rm sp}(\rho)=\mathop{\rm Min}_{\bm{c}}\,I^{\bm {c}}(\rho)=
\mathop{\rm Min}_{\{c_k\}}\sum_k I^{c_k}(\rho)
\,,\label{iloss}
\end{equation}
measures then the minimum loss of information due to such type of measurement.
This \emph{one-body information loss} clearly satisfies $I^{\rm sp}(\rho)\geq
0$, since it is a sum of non-negative terms, with  $I^{\rm sp}(\rho)=0$ iff
fermion operators $c_k$ exist such that $\rho'(k)=\rho$ $\forall$ $k$, i.e.,
iff
\begin{equation}[\rho,c^\dagger_k c_k]=0\,,\;\;k=1,\ldots,n,\label{ckd}\end{equation}
which occurs iff $\rho$ is diagonal in a set of Slater determinants
$\{\prod_k (c^\dagger_k)^{n_k}|0\rangle\}$
in the same sp basis (the common eigenvectors of all $c^\dagger_k c_k$). Such states
include the typical uncorrelated thermal-like states $\rho\propto\exp[-\beta
H]$, with $H=\sum_k \varepsilon_k c^\dagger_k c_k$, but also any convex
combination of Slater determinants in the same basis. These combinations play
here the role of ``classically'' correlated states.

Note that if (\ref{ckd}) holds, for $k\neq l$ we have $\langle c^\dagger_k
c_l\rangle=\langle [c^\dagger_k c_k ,c^\dagger_k c_l]\rangle=0$, so that the
operators $c_k$  diagonalize the sp density matrix: $\langle c^\dagger_k
c_l\rangle=\lambda_k \delta_{kl}$. Therefore, even though the operators
minimizing $I^{\bm c}(\rho)$ may not diagonalize $\rho^{\rm sp}$ in general,
they will if $I^{\rm sp}(\rho)=0$. Thus, $I^{\bm c}(\rho)>0$ in all sp bases
diagonalizing $\rho^{\rm sp}$ imply $I^{\rm sp}(\rho)>0$.

For a pure state $\rho=|\psi\rangle\langle\psi|$ equation (\ref{Ipk}) implies
\begin{equation}
I^{\bm{c}}(|\psi\rangle)=\sum_k S(k)=\sum_k h(p_k)\,,
\end{equation}
which is just a concave function of the diagonal elements of $\rho^{\rm sp}$.
Therefore, in this case its minimum over all bases is obtained when $p_k$ are
its eigenvalues, i.e., when the $c_k$'s are the fermion operators diagonalizing
$\rho^{\rm sp}$:
\begin{equation}
I^{\rm sp}(|\psi\rangle)={\rm tr}\,h(\rho^{\rm sp})=S^{\rm sp}(|\psi\rangle)\,.\label{Ispp}
\end{equation}
This result coincides with the entanglement entropy defined in the previous
section, in analogy with the information loss for the distinguishable case
\cite{RCC.10}, which also coincides with the entanglement entropy for pure
states.

On the other hand, it is seen from (\ref{spk}) that for a general mixed state
$\rho$, $S(\rho'(k))\geq S(k)=h(p_k)$  and hence,
\begin{equation}
\mathop{\rm Min}_{\{c_k\}}\sum_k S(\rho'(k))\geq {\rm tr}\,h(\rho^{\rm sp})\,.
\end{equation}

As in the case of the one-body entanglement entropy, if quasiparticle mode
measurements are to be allowed then the minimization extends to all
quasiparticle bases, i.e., sets of operators $\{a_k, a^\dagger_k\}$ related to
the original fermion operators $\{c_i, c^\dagger_i\}$ through a Bogoliubov
transformation (\ref{Bg}). It can be seen by repeating the argument used above
that this information loss,
\begin{equation}
I^{\rm qsp}(\rho)= \mathop{\rm Min}_{\{a_k\}}\sum_k
 [S(\rho'(k))-S(\rho)]\label{Sgloss}\end{equation}
satisfies $I^{\rm qsp}(\rho)\geq 0$, with equality iff quasiparticle operators
$a_k$ exist such that
\begin{equation}
[\rho,a^\dagger_k a_k]=0\,,\;\;k=1,\ldots,n\,,\label{akd}
\end{equation}
i.e., iff $\rho$ is diagonal in a set of  {\it quasiparticle} Slater
determinants in the same basis (common eigenvectors of all $a^\dagger_k a_k$).
Moreover, as this extended minimization includes the previous one as a
particular case, we have
\begin{equation}
I^{\rm qsp}(\rho)\leq I^{\rm sp}(\rho)\,.
\end{equation}
For pure states, the minimum $I^{\rm qsp}$ is obtained for those $a^\dagger_k$
diagonalizing the extended density matrix (\ref{qsp}), which in analogy with
(\ref{Ispp}) yields
\begin{equation}
I^{\rm qsp}(|\psi\rangle)=-{\rm tr}'\rho^{\rm qsp}\log_2\rho^{\rm qsp}=S^{\rm qsp}(|\psi\rangle)\,,
\end{equation}
where $S^{\rm qsp}$ is the generalized entanglement entropy (\ref{sqsp2}). As
in the previous case, if (\ref{akd}) holds, the operators $a_k$ diagonalize
$\rho^{\rm qsp}$, implying that even though such operators may not minimize
$I^{\bm a}(\rho)$ in general,  they will if $I^{\rm qsp}(\rho)=0$.

Let us also remark that $I^{\rm sp}(\rho)$ remains invariant under standard unitary
one-body transformations
\begin{equation}
\rho\rightarrow U\rho U^\dagger\,,\;\;U=\exp[-\imath t\sum_{i,j} h_{ij}c^\dagger_i c_j]\,,
\end{equation}
with $h_{ij}=h_{ji}^*$, since they just imply a unitary transformation of the
fermion operators $\bm{c}=(c_1,\ldots,c_n)^T$:
\begin{equation}
\bm{c}\rightarrow U\bm{c}U^\dagger=\exp[\imath th]\bm{c},
\end{equation}
with $\rho^{\rm sp}\rightarrow \exp[\imath t h]\rho^{\rm sp}\exp[-\imath t h]$.
These transformations map Slater determinants onto Slater determinants.
Similarly, $I^{\rm qsp}(\rho)$ remains invariant under general
 one-body transformations
\begin{eqnarray}
\rho&\rightarrow& W\rho W^\dagger\,,\;\;W=\exp[-\imath tH]\,,\nonumber\\
H&=&\sum_{i,j}h_{ij}c^\dagger_i c_j+{\textstyle\frac{1}{2}}\Delta_{ij}
(c^\dagger_i c^\dagger_j+c_j c_i)\label{H1}\\
&=&{\textstyle\frac{1}{2}(\bm{c}^\dagger\;\bm{c}){\cal H}\left(\begin{array}{c}\bm{c}\\
\bm{c}^\dagger\end{array}\right)},\;\;
{\cal H}=\left(\begin{array}{cc}h&\Delta\\-\Delta^*&-h^*\end{array}\right)\,,\label{H2}
\end{eqnarray}
where $\Delta^T=-\Delta$, since they just imply a Bogoliubov transformation of
the fermion operators:
\[\left(\begin{array}{c}\bm{c}\\\bm{c}^\dagger\end{array}\right)\rightarrow
 \exp[\imath t{\cal H}]\left(\begin{array}{c}\bm{c}\\\bm{c}^\dagger\end{array}\right)\]
with $\rho^{\rm qsp}\rightarrow \exp[\imath t{\cal H}]\rho^{\rm
qsp}\exp[-\imath t {\cal H}]$. Therefore, evolution under Hamiltonians $H$ of the
previous form will not alter the value of $I^{\rm qsp}(\rho)$.

Finally, it is worth remarking that
\begin{equation}I^{\rm qsp}(\rho)=0\,,\end{equation}
for {\it any} state $\rho$ with support in a {\it two-dimensional} sp space
($n=2$) which commutes  with the parity $P$.
\\{\it Proof}: Let
\begin{equation}\rho=\sum_{\nu=1,2} q_\nu^-|\psi^\nu_-\rangle\langle\psi^\nu_-|
+q_\nu^+|\psi^\nu_+\rangle\langle\psi^\nu_+|\label{rho2}\end{equation}
be the spectral decomposition of such $\rho$, with
\[|\psi^\nu_-\rangle=(\alpha_1^\nu c^\dagger_1+\alpha^\nu_2c^\dagger_2)|0\rangle\,,\;\;
|\psi^\nu_+\rangle=(\beta^\nu_1+\beta^\nu_2
 c^\dagger_2c^\dagger_1)|0\rangle\,,\]
its odd and even parity orthogonal eigenstates
($\alpha^2_j=(-)^j\bar{\alpha}^1_{3-j}$, $\beta^2_j=(-)^j\bar{\beta}^1_{3-j}$).
It is easily seen that these states can be all  written as $0$, $1$ or
$2$-quasiparticle states in a common basis. In fact, $\langle
c^\dagger_2c_1\rangle=(q_1^--q_2^-)\bar{\alpha}^1_2\alpha^1_1$, $\langle
c^\dagger_2c^\dagger_1\rangle=(q_1^+-q_2^+)\bar{\beta}^1_2\beta^1_1$. Hence, in
terms of the quasiparticle operators $a_k$ that diagonalize $\rho^{\rm qsp}$
($\langle a^\dagger_2
a_1\rangle=(q_1^--q_2^-){\bar{\alpha'}}^1_2{{\alpha}'}^1_1=0$, $\langle
a^\dagger_2a^\dagger_1\rangle=(q_1^+-q_2^+){\bar{\beta'}}^1_2{{\beta}'}^1_1=0$),
necessarily $|\psi^1_-\rangle\propto a^\dagger_1|0'\rangle$,
$|\psi^2_-\rangle\propto a^\dagger_2|0'\rangle$,
$|\psi^1_+\rangle\propto|0'\rangle$ and $|\psi^2_+\rangle\propto
a^\dagger_2a^\dagger_1|0'\rangle$ if $q_1^\pm\neq q_2^\pm$, with $|0'\rangle$
the vacuum of these quasiparticles (of the same parity as $|0\rangle$). And in
case of degeneracy we may always choose the eigenstates
$|\psi^\nu_{\pm}\rangle$ of the previous forms. Hence $\rho$  is always
diagonal in a quasiparticle Slater determinant basis, implying $I^{\rm
qsp}(\rho)=0$.

Note, however, that $I^{\rm sp}(\rho)>0$ unless $[\rho,N]=0$, i.e.,
$\langle c^\dagger_2c^\dagger_1\rangle=0$, and that $\rho$ is not necessarily
of the uncorrelated form
$\propto\exp[-\sum_{\nu=1,2}\varepsilon_\nu a^\dagger_\nu a_\nu]$ unless
$\langle a^\dagger_1a_1a^\dagger_2a_2\rangle=\langle
a^\dagger_1a_1\rangle\langle a^\dagger_2a_2\rangle$, i.e.,
$q^+_2=(q_1^-+q_2^+)(q_2^-+q_2^+)$. \qed

On the other hand, if the support is a three dimensional sp space ($n=3$), then
$I^{\rm qsp}(\rho)=0$ for any $\rho$ of {\it definite} parity (but not for any
$\rho$ commuting with $P$).
\\{\it Proof}: Let $\rho=\sum_{\nu=1}^4 q_\nu|\psi^\nu\rangle\langle\psi^\nu|$
be the spectral decomposition of $\rho$. If its eigenstates have all odd parity, i.e.
\[|\psi^\nu\rangle=(\sum_{j=1}^3\alpha^\nu_j c_j^\dagger+
\alpha^\nu_4 c^{\dagger}_{3}
 c^{\dagger}_{2}c^{\dagger}_{1})|0\rangle\,,\;\;\nu=1,\ldots,4\,,\]
then $\langle c^\dagger_jc_k\rangle=\sum_\nu
q_\nu(\bar{\alpha}^\nu_j\alpha^\nu_k+\delta_{jk}|\alpha^\nu_4|^2$), $\langle
c^\dagger_jc^\dagger_k\rangle=-\sum_{\nu,l}
q_\nu\epsilon_{jkl}\bar{\alpha}^\nu_4\alpha^\nu_l$, with $\epsilon_{jkl}$ the
fully antisymmetric tensor. Hence, if expressed in terms of the quasiparticle
operators $a_k$ that diagonalize $\rho^{\rm qsp}$ ($c^\dagger_j\rightarrow
a^\dagger_j$, $|0\rangle\rightarrow |0'\rangle$, $\alpha^\nu_j\rightarrow
{\alpha'}^\nu_j$, with $\langle a^\dagger_j a_k\rangle=f_j\delta_{jk}$,
$\langle a^\dagger_j a^\dagger_k\rangle=0$), the previous relations together
with $\langle\psi^\nu|\psi^{\nu'}\rangle=\sum_{\mu=1}^4
{\bar{\alpha'}}^\nu_\mu{\alpha'}^{\nu'}_\mu=\delta^{\nu\nu'}$ imply, for distinct
$q^\nu$'s, ${\alpha'}^\nu_\mu\propto \delta^\nu_\mu$, i.e. $|\psi^\nu\rangle\propto
a^\dagger_\nu|0'\rangle$, $\nu=1,2,3$ and $|\psi^4\rangle \propto a^\dagger_3
a^\dagger_2 a^\dagger_1|0'\rangle$. Hence $\rho$ is diagonal in a quasiparticle
Slater determinant basis and $I^{\rm qsp}(\rho)=0$ (but $I^{\rm sp}(\rho)\neq
0$ if $[\rho,N]\neq 0$). The same procedure can be applied for an even parity
$\rho$ (which can be recast in the previous odd-parity form after a
particle-hole transformation $c^\dagger_i \rightarrow c_i$ of one of the
operators).

Note, however, that if $\rho$ contains eigenstates of different parity, $I^{\rm
qsp}(\rho)$ can be positive since the fermion quasiparticle operators of the
normal form for each parity will not coincide in general. \qed

\subsection{General stationary condition}
Let us now derive  the general stationary equations that must be satisfied by
the set of operators $\{a_k\}$ minimizing the generalized one-body
information loss (\ref{Sgloss}). After a measurement of the occupancy of a
corresponding level $k$, the ensuing state $\rho'(k)$, Eq.\ (\ref{rhopk}),
has eigenstates $|\phi^\mu_{k(\bar{k})}\rangle$ with eigenvalues
\[q^\mu_{k(\bar{k})}=\langle\phi^\mu_{k(\bar{k})}|\rho(k)|\phi^\mu_{k(\bar{k})}\rangle
 =\langle\phi^\mu_{k(\bar{k})}|\rho|\phi^\mu_{k(\bar{k})}\rangle\,.\]
Consider now a small variation of the measurement basis determined by a general
one-body transformation
\[W=\exp[-\imath\epsilon H]\approx 1-\imath\epsilon H\,,\]
with $H$ of the general form (\ref{H1})--(\ref{H2}).
 We then have
$\delta q^\mu_{k(\bar{k})}=-\imath\epsilon \langle
\phi^\mu_{k(\bar{k})}|[\rho,H]|\phi^\mu_{k(\bar{k})}\rangle$ up to lowest order
in $\epsilon$, which implies, for the information loss (\ref{Sgloss}),
\begin{eqnarray}
\delta I^{\rm qsp}(\rho)&=&\sum_{k,\mu} f'(q^\mu_k)\delta q^\mu_k
+f'(q^\mu_{\bar{k}})\delta q^\mu_{\bar{k}}\nonumber\\
&=&-\imath\epsilon{\rm Tr}\,\left([\sum_k
 f'(\rho'(k)),\rho]H\right)\,,\label{statc}\end{eqnarray}
where $f'(\rho)=-\log_2 \rho$. The condition $\delta I^{\rm qsp}(\rho)=0$
for {\it arbitrary} $H$ then leads to the stationary equations
${\rm Tr}\,[\sum_k f'(\rho'(k)),\rho]c^\dagger_i c_j=0$,
${\rm Tr}\,[\sum_k f'(\rho'(k)),\rho]c^\dagger_i c^\dagger_j=0$ 
$\forall$ $i,j$, which reduce to
\begin{eqnarray}
 {\rm Tr}\,\rho\,[f'(\rho'(k))+f'(\rho'(l)),a^\dagger_k a_l]&=&0\,,\label{st1}\\
 {\rm Tr}\,\rho\,[f'(\rho'(k))+f'(\rho'(l)),a^\dagger_k a^\dagger_l]&=&0\,,\label{st2}
\end{eqnarray}
$\forall$  $k\neq l$ when expressed in terms of the quasiparticle operators
determining the measurement basis ($[\rho'(k),a^\dagger_k a_k]=0$ $\forall$ $k$),
 since ${\rm Tr}\,[f'(\rho(k)),\rho]a^\dagger_j a_l=0$
if $j\neq k\neq l$,  and also if $j=l=k$.
In the case of $I^{\rm sp}(\rho)$, just eq.\ (\ref{st1}) is to be considered ($\Delta=0$ in $H$).

For a pure state $\rho=|\psi\rangle\langle \psi|$,
Eqs.\ (\ref{st1})--(\ref{st2}) become
\begin{eqnarray}
[g(p_k)-g(p_{\bar{k}})-g(p_l)+g(p_{\bar{l}})]\langle \psi|a^\dagger_k a_l|\psi\rangle&=&0
\,,\label{st11}\\
{[g(p_k)-g(p_{\bar{k}})+g(p_l)-g(p_{\bar{l}})]}\langle \psi|a^\dagger_k a^\dagger_l|\psi\rangle
&=&0\,,\label{st22}
\end{eqnarray}
for $k\neq l$, where $g(p)=f'(p)$ and $p_{k(\bar{k})}$ are the probabilities
of finding state $k$ occupied (empty) in $|\psi\rangle$.  It is then verified
that they are fulfilled by those $a_k$ diagonalizing
the extended density matrix (\ref{qsp})
($\langle a^\dagger_k a_l\rangle=\langle a^\dagger_k a^\dagger_l\rangle=0$).

\subsection{Generalized one-body information loss}

We may directly extend all previous considerations to more general entropic
forms. We first consider the generalized trace form entropies
\cite{RC.02,CR.02}
\begin{equation}S_f(\rho)={\rm Tr}\,f(\rho)\,,\label{Sf}\end{equation}
where $f:[0,1]\rightarrow\mathbb{R}$ is a smooth strictly concave real function
satisfying $f(0)=f(1)=0$. For $f(\rho)=-\rho\log_2 \rho$, $S_f(\rho)$ becomes
the von Neumann entropy $S(\rho)$, whereas for $f(\rho)=2\rho(1-\rho)$, it
becomes the quadratic entropy $S_2(\rho)$, which is just a decreasing function
of the purity ${\rm Tr}\,\rho^2$ and corresponds to the linear approximation
$-\rho\ln\rho\approx\rho(1-\rho)$. It  does not  require the explicit
knowledge of the eigenvalues of $\rho$, being then easier to determine than
$S(\rho)$ \cite{GR.14,NK.13}. Moreover, the associated generalization of the
one-body entanglement entropy  (\ref{sqsp2}), $S_f^{\rm qsp}(|\psi\rangle)={\rm
tr'}f({\rho'}^{\rm qsp})$  \cite{GR.15}, becomes
\begin{equation}
S_2^{\rm qsp}(|\psi\rangle)=2\,{\rm tr'}\rho^{\rm qsp} (1-\rho^{\rm qsp})
 =4\sum_k f_k(1-f_k)\,, \label{sqqsp}\end{equation}
which is just the sum of the {\it fluctuations} of the occupancies of all
single quasiparticle levels. More generally, these entropies include the family
of Tsallis entropies \cite{TS.09}, obtained for
$f(\rho)=(\rho-\rho^q)/(1-2^{1-q})$ with $q>0$, $q\neq 1$,  which become
proportional to the von Neumann  and quadratic entropies for $q\rightarrow 1$
and $q=2$ respectively. We use here the normalization ${\rm Tr}f(\rho)=1$ for a
maximally mixed single qubit state $\rho$.

A function $f$ defined with the properties stated above ensures that all entropies $S_f(\rho)$
 satisfy \cite{CR.02,RC.02} i) $S_f(\rho)\geq 0\,\forall \rho$, with
$S_f(\rho)=0$ iff $\rho^2=\rho$, ii) $S_f(\sum_i q_i\rho_i)\geq\sum_iq_iS_f(\rho_i)$
for $q_i\geq 0$, $\sum_i q_i=1$ (concavity) and iii),
\begin{equation}
\rho'\prec\rho\Rightarrow S_f(\rho')\geq S_f(\rho)\,,
 \label{Smaj}\end{equation}
 where $\rho'\prec\rho$ indicates that the sorted set $\{q'_i\}$ of eigenvalues of $\rho'$
  ($q'_i>q'_j$ if $i<j$) is {\it majorized} \cite{Bh.97,RC.02,RCC.10} by the
  sorted set $\{q_i\}$ of eigenvalues of $\rho$:
\[
\rho'\prec\rho\Leftrightarrow\sum_{i=1}^k q'_i\leq\sum_{i=1}^k q_i\,,\;\; k=1,\ldots,d
\]
with $\sum_{i=1}^d q'_i=\sum_{i=1}^d q_i$. Therefore, $S_f(\rho)$ increases with
increasing mixedness of $\rho$.

This last property allows the straightforward extension of the results of the
previous section to the present more general entropic forms.  Indeed,
$\rho'(k)$ in (\ref{rhopk}) is majorized by the original state $\rho$,
$\rho'(k)\prec\rho$, implying $S_f(\rho'(k))\geq S_f(\rho)$ and hence
\begin{equation}
I_f^{c_k}(\rho)\equiv S_f(\rho'(k))-S_f(\rho)\geq 0,\label{If}
\end{equation}
with $I_f^{c_k}(\rho)=0$ iff $\rho'(k)=\rho$. Eq.\ (\ref{If}) is a measure of
the information loss due to the unread measurement of the occupation of sp
state $k$, and is also an indicator of the entanglement generated between
the system and measurement device, according to the generalized entropic
separability criterion \cite{RC.02}. By summing over all states in a basis
and minimizing over all possible bases of $\mathcal{S}$ we obtain
\begin{equation}
I_f^{\rm sp}(\rho)=\mathop{\rm Min}_{\{c_k\}}\sum_k I_f^{c_k}
=\mathop{\rm Min}_{\{c_k\}}\sum_k S_f(\rho'(k))-S_f(\rho) \label{giloss}\,,
\end{equation}
which satisfies $I_f^{\rm sp}(\rho)\geq 0$, with $I_f^{\rm sp}(\rho)=0$ iff
$\rho$ is diagonal in a set of Slater Determinants in the same sp basis
($[\rho,c^\dagger_k c_k]=0$ $\forall$ $k$).  The minimization may again be
extended to all quasiparticle bases through a Bogoliubov transformation,
leading to the quantity
\begin{equation}I_f^{\rm qsp}(\rho)=\mathop{\rm Min}_{\{a_k\}}\sum_k S_f(\rho'(k))-S_f(\rho)\,,
\label{Ifqsp}
\end{equation}
which satisfies $0\leq I_f^{\rm  qsp}(\rho)\leq I_f^{\rm sp}(\rho)$, with
$I_f^{\rm qsp}(\rho)=0$ iff $\rho$ is diagonal in a basis of quasiparticle
Slater determinants ($[\rho,a^\dagger_k a_k]=0$ $\forall$ $k$). For pure states
$\rho^2=\rho$,  $I_f^{\rm qsp}(\rho)=S_f(\rho^{\rm qsp})$  becomes the
generalized entropy of the extended one-body density matrix. Let us remark that
the general stationary conditions (\ref{st1})--(\ref{st22})  remain valid for
the present generalization, with $f'$ denoting now the derivative of the
entropic function $f$.

Eq.\ (\ref{Smaj}) remains also valid for Schur-concave functions of $\rho$
\cite{Bh.97}, which include, in particular, increasing functions of the
previous entropies $S_f(\rho)$. An  example is provided by the quantum version
of the Renyi entropies  \cite{BS.93},
\begin{equation}
S^R_{q}(\rho)=\frac{\log_2({\rm Tr}\,
\rho^q)}{1-q}=\frac{\log_2[1-(1-2^{1-q})S_q(\rho)]}{1-q}\,,\label{renyi}
\end{equation}
where $q>0$, $q\neq 1$, which are just increasing functions of the Tsallis
entropies $S_q(\rho)$ and approach the von Neumann entropy for $q\rightarrow
1$. The definition of information loss straightforwardly extends to these
entropies. The logarithm in (\ref{renyi}) implies additivity, i.e.,
$S^R_q(\rho\otimes\sigma)=S^R_q(\rho)+S^R_q(\sigma)$, which ensures that the
addition of an uncorrelated ancilla to the system
($\rho\rightarrow\rho\otimes\sigma$) has no effect  on the associated
information deficit \cite{CCR.15} $I_{R_q}(\rho)=\sum_k
S^R_q(\rho'(k))-S^R_q(\rho)$. Nonetheless, the optimization problem for
$I_{R_q}(\rho)$ is the same as that for $I_q(\rho)=\sum_k
S_q(\rho'(k))-S_q(\rho)$.

We finally mention that for fermions we may consider yet another way of adding
an ancilla to our system $S$, by expanding its sp Hilbert space
$\mathcal{S}\rightarrow\mathcal{S}\oplus\mathcal{A}$. A non correlated state of
the system $S+A$ will then have the form $\rho_S\rho_A$, where $\rho_A$ and
$\rho_B$ involve creation and annihilation operators of single particle states
in $\mathcal{S}$ and $\mathcal{A}$ respectively (see Appendix). We have ${\rm
Tr}\,\rho_{S}\rho_A ={\rm Tr}\,\rho_{S}\,{\rm Tr}\rho_{A}$ if traces are taken
in a grand canonical ensemble (as we are here assuming), and hence
$S^R_q(\rho_{S}\rho_A)=S^R_q(\rho_S)+S^R_q(\rho_A)$.

\section{Application}
\subsection{Mixture of pure state plus maximally mixed state}
Let us now consider  the mixture
\begin{equation}
\rho= w|\psi\rangle\langle\psi|+\frac{1-w}{d}I_d\label{W},
\end{equation}
with $0\leq w\leq 1$ and $d$ the dimension of the state space. After an unread
measurement of mode $|k\rangle$ the state reads
\begin{equation}
\rho'(k)=w(p_k|\psi_k\rangle\langle \psi_k|+p_{\bar k}|\psi_{\bar k}\rangle\langle \psi_{\bar k}|)+
\frac{1-w}{d}I_d,\label{rhkk}
\end{equation}
with $p_{k(\bar{k})}=\langle c^\dagger_k c_k\rangle$ ($\langle c_k c^\dagger_k\rangle$)
the probability of finding mode $k$ occupied (empty) in $|\psi\rangle$.
Its eigenvalues are $q_{k(\bar{k})}=wp_{k(\bar{k})}+\frac{1-w}{d}$ and
 $\frac{1-w}{d}$, so that
\begin{eqnarray}
S_f(\rho'(k))&=&f(q_k)+f(q_{\bar{k}})+(d-2)f({\textstyle\frac{1-w}{d}})\,,\\
I_f^{\bm{c}}(\rho)&=&\sum_k[S(\rho'(k))-S(\rho)]\nonumber\\
&=&\sum_k [f(q_k)+f(q_{\bar{k}})-f(w+{\textstyle\frac{1-w}{d}})-f({\textstyle\frac{1-w}{d}})]\,.
\label{eqr}
\end{eqnarray}
We now show that the minimum of $I_f^{\bm c}(\rho)$ over all sp basis of
$\mathcal{S}$ is reached for the operators $\{c'_k\}$ that diagonalize the
one-body density matrix $\rho^{\rm sp}_{ij}=\langle c^\dagger_j c_i\rangle$,
while the minimum over all quasiparticle basis is attained for those  $\{a_k\}$
diagonalizing the corresponding extended one-body density matrix $\rho^{\rm
qsp}$.
\begin{proof}
Denoting with $\{\lambda_k=\langle {c'}_k^\dagger c'_k\rangle\}$ the set of
eigenvalues of the one-body density matrix $\rho^{\rm sp}$, such that $\langle
{c'}_k^\dagger {c'}_j\rangle =\lambda_k\delta_{kj}$, this distribution
majorizes any other diagonal of the matrix, implying $\{p_k=\langle
{c}_k^\dagger c_k\rangle\}\prec\{\lambda_k\}$ for the sorted sets. Hence,
$\{q_k\}\prec\{{q'}_k\}$ if $q'_k=w\lambda_k+\frac{1-w}{d}$ and $0\leq w\leq
1$, and also  $\{q_{\bar{k}}\}\prec\{q'_{\bar{k}}\}$ if
$q'_{\bar{k}}=w\lambda_{\bar{k}}+\frac{1-w}{d}$, since
$\{p_k\}\prec\{\lambda_k\}$ implies
$\{p_{\bar{k}}=1-p_k\}\prec\{\lambda_{\bar{k}}=1-\lambda_k\}$. Therefore, Eq.\
(\ref{Smaj})
 leads to
\begin{equation}
\sum_k f(q_k)+f(q_{\bar{k}})\geq\sum_k f({q'}_{k})+f({q'}_{\bar{k}}),\label{ineq}
\end{equation}
implying $I_f^{\rm sp}(\rho)={\rm
Min}_{\bm{c}}\,I_f^{\bm{c}}(\rho)=I_f^{\bm{c}'}(\rho)$. This result also
follows directly from the concavity of $f$ and the relation
$q_{k(\bar{k})}=\sum_{k'}|U_{kk'}|^2 {q'}_{k'(\bar{k}')}$, with $U$ the unitary
matrix diagonalizing $\rho^{\rm sp}$.

When quasiparticles are also considered, we note that both $p_k$ and
$p_{\bar{k}}=1-p_k$ are diagonal elements of $\rho^{\rm qsp}$, so that the
enlarged sorted set $\{p_k,p_{\bar{k}}\}$ is majorized by the whole sorted set
$\{f_k=\langle a^\dagger_k a_k\rangle, f_{\bar{k}}=1-f_k\}$ of eigenvalues of
$\rho^{\rm qsp}$ ($\langle a^\dagger_k a_l\rangle=\delta_{kl}f_k$, $\langle
a^\dagger_k a^\dagger_l\rangle=0$). Therefore, $\{q_{k},q_{\bar{k}}\}\prec \{
{q'}_{k},{q'}_{\bar{k}}\}$ for ${q'}_{k(\bar{k})}=w
f_{k(\bar{k})}+\frac{1-w}{d}$, implying Eq.\ (\ref{ineq}) and hence $I_f^{\rm
qsp}(\rho)=I_f^{\bm{a}}(\rho)$. This result also follows from the relation
$q_{k}=\sum_{k'}|W_{kk'}|^2 {q'}_{k'}$ between the elements of the enlarged
sets, with $W$ the matrix diagonalizing $\rho^{\rm qsp}$.
\end{proof}

These results are valid  for the von Neumann entropy as well as for the
generalized entropic forms $S_f$, and are evidently in agreement with the
stationary conditions (\ref{st1})--(\ref{st2}), since for the state (\ref{W})
they become proportional to Eqs.\ (\ref{st11})--(\ref{st22}). They also hold if
$\rho$ has a definite parity $P$, i.e., if $I_d$ stands for the projector onto
the same parity as that of $|\psi\rangle$,  in which case $d\rightarrow d/2$ in
(\ref{W})--(\ref{ineq}). And if $|\psi\rangle$ has definite fermion number $N$,
they are also valid in a canonical ensemble, with $d\rightarrow (^{\,n}_N)$.

The previous arguments also imply that if  $|\psi'\rangle$ is a state whose one
body density matrix is majorized by that of $|\psi\rangle$, ${\rho'}^{\rm
qsp}\prec\rho^{\rm qsp}$, and $\rho'=w|\psi'\rangle\langle\psi'|+(1-w)I_d/d$,
then
\begin{equation} I^{\rm qsp}_f(\rho')\geq I^{\rm qsp}_f(\rho)\,,\end{equation}
$\forall\; w\in [0,1]$ and $\forall$ $S_f$.  This general inequality
reflects the rigorously stronger entanglement of $|\psi'\rangle$, in the sense
that $S_f^{\rm qsp}(|\psi'\rangle)=S_f({\rho'}^{\rm qsp})\geq S_f(\rho^{\rm
qsp})$ for {\it all} entropies $S_f$ if ${\rho'}^{\rm qsp}\prec\rho^{\rm qsp}$,
and indicates that the value of $I_f^{\rm qsp}$ in the mixture (\ref{W}) is
indeed driven by the entanglement of the pure state. The same relation holds
for $I_f^{\rm sp}$ if ${\rho'}^{\rm sp}\prec \rho^{\rm sp}$.

Let us  point out that $I_f^{\rm qsp}(\rho)$ is a strictly increasing function
of $w$ for any concave $f$, with  $I_f^{\rm qsp}(\rho)>0$ for any $w>0$. From
Eq.\ (\ref{eqr}) it is seen that it  exhibits for small $\omega$ a {\it
universal} initial quadratic increase,  given by
\begin{equation} I_f^{\rm qsp}(\rho)\approx w^2|f''(d^{-1})|\sum_k f_k(1-f_k)
\,,\label{quad}\end{equation}
with $f_k$ the eigenvalues of the $\rho^{\rm
qsp}$ determined by $|\psi\rangle$, which is just proportional to the quadratic
entanglement of $|\psi\rangle$, Eq.\ (\ref{sqqsp}). For $I_f^{\rm sp}(\rho)$ we
should just replace $f_k$ by the eigenvalues $\lambda_k$ of $\rho^{\rm sp}$. In
the case of the quadratic entropy $S_2$,  Eq.\ (\ref{quad}) is of course {\it
exact} $\forall$ $w\in[0,1]$ and independent of $d$ ($|f''(d^{-1})|=4$).

Previous results are  then similar to those obtained for distinguishable
bipartite quantum systems \cite{RCC.10}, where the role played here by the
basis diagonalizing $\rho^{\rm qsp}$ corresponds there to the local part of the
Schmidt basis of the pertinent pure state.

\subsection{The case of four single particle levels}

We now focus on the special case of a fermion system with $n={\rm
dim}(\mathcal{S})=4$, where the entanglement of formation for general states
can be analytically evaluated \cite{Sch.01,GR.15}. For simplicity we will
consider mixed states with definite parity, which will be choose as odd. A
general pure state will be then a linear combination of single fermion states
and three fermion states. Therefore, it can be written as
\begin{equation}
|\psi\rangle=\sum_i^4(\alpha_i c^\dagger_i|0\rangle+{\bar\beta}_ic_i|{\bar 0}\rangle),
\label{st3}
\end{equation}
where $|{\bar 0}\rangle=c^\dagger_1c^\dagger_2c^\dagger_3c^\dagger_4|0\rangle$
is the fully occupied state and $\bm\alpha, \bm\beta$ are four-dimensional
complex vectors satisfying $|\bm\alpha|^2+|\bm\beta|^2=1$. It can be shown
\cite{GR.15} that the eigenvalues of the generalized one-body density matrix
$\rho^{\rm qsp}$ of such state are four-fold degenerate and given by
\begin{equation}
f_{\pm}=\frac{1\pm\sqrt{1-C^2(|\psi\rangle)}}{2},
\end{equation}
where $C(|\psi\rangle)=2|\bm\beta^\dagger\bm\alpha|$ is the \emph{generalized
Slater correlation measure}, satisfying $0\leq C\leq 1$. This result implies
that there is always a quasiparticle basis in which the state (\ref{st3}) takes
the normal form
\begin{equation}|\psi\rangle=(\sqrt{f_+}\,a_1^\dagger+\sqrt{1-f_+}
\,a_2^\dagger a_3^\dagger a_4^\dagger)|0\rangle
\,,\end{equation}
with $|0\rangle$ denoting now the vacuum of the operators $a_i$.  In terms of $C$ the entanglement
entropy (\ref{sqsp2}) becomes then
\begin{equation}
S^{\rm qsp}(|\psi\rangle)=4h \left ({\textstyle\frac{1+\sqrt{1-C^2(|\psi\rangle)}}{2}}\right),
\end{equation}
so $C(|\psi\rangle)$ plays the role of a fermionic concurrence. As in the
two-qubit case, for a mixed state $\rho$ the convex roof extension of $S^{\rm
qsp}$ can be similarly evaluated as
\begin{equation}
S^{\rm qsp}(\rho)=4h\left({\textstyle\frac{1+\sqrt{1-C^2(\rho)}}{2}}\right),\label{cr}
\end{equation}
where $C(\rho)$ is the convex roof extension of the pure state concurrence
defined above \cite{GR.15}. This quantity vanishes iff $\rho$ can be written as
a convex combination of particle or quasiparticle Slater determinants.

Let us now consider the state (\ref{W}) with a maximally entangled state
$|\psi\rangle=\frac{1}{\sqrt{2}}(a^\dagger_1+a^\dagger_2a^\dagger_3a^\dagger_4)|0\rangle$,
which leads to $f_{\pm}=1/2$. The fermionic concurrence is then $C(\rho)={\rm
Max}[(7w-3)/4, 0]$ \cite{GR.14} and therefore, Eq.\ (\ref{cr}) leads to
\begin{equation}
S^{\rm qsp}(\rho)=\left\{\begin{array}{ll}0&w\leq 3/7\\4h\left(\frac{4+\sqrt{7[1+w(6-7w)]}}{8}\right)
&w>3/7\end{array}\right. \,.
\end{equation}
The information loss $I_f^{\rm qsp}(\rho)$ can be easily evaluated from Eq.\ (\ref{eqr}),
since all single particle levels have probability $1/2$ of being occupied:
\begin{equation}
I^{\rm qsp}_f(\rho)=4[{\textstyle 2f(\frac{3w+1}{8})-f(\frac{7w+1}{8})-f(\frac{1-w}{8})}]\,.
\end{equation}

\begin{figure}[htp]
\centering
\includegraphics[scale=0.65]{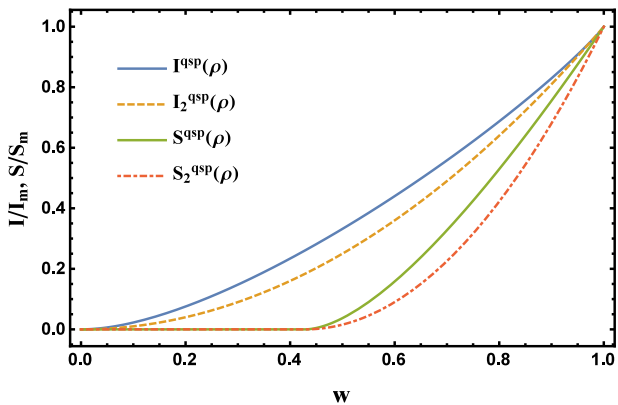}
\caption{Quadratic (dashed line) and von Neumann (solid line) information loss
$I^{\rm qsp}(\rho)$ and entanglement of formation $S^{\rm qsp}(\rho)$
(normalized to their maximum values), as a function of $w$ for  the mixture
(\ref{W}) with $n=4$ and  definite odd parity, for a maximally entangled state
$|\psi\rangle$.} \label{1}
\end{figure}

Fig.\ \ref{1} depicts  this information loss in the von Neumann case
($f(p)=-p\log_2 p$) and in the quadratic case ($f(p)=2p(1-p)$), together with
the corresponding entanglement of formations $S^{\rm qsp}(\rho)$ and $S_2^{\rm
qsp}(\rho)=4C^2(\rho)$,  as a function of $w$. While there is a threshold
value, $w=3/7$, below which the state remains separable, $I_f^{\rm
qsp}(\rho)>0$ as soon as $\rho$ departs from the maximally mixed state, as
given by Eq.\ (\ref{quad}), implying that the information loss detects ``non
classical'' correlations beyond entanglement. In addition, it is worth noting
that while  $I_2^{\rm qsp}(\rho)$ is an upper bound to $S_2^{\rm qsp}(\rho)$
for the present states $\forall$ $w\in(0,1)$, this is not strictly the case for
the von Neumann based  quantities, as $I^{\rm qsp}(\rho)<S^{\rm qsp}(\rho)$ for
$w$ below but very close to $1$.

\subsection{Two fermion states}

We now assume $|\psi\rangle$ in (\ref{W}) is a two-fermion state. These states are  of the form
\begin{equation}
|\psi\rangle=\frac{1}{2}\sum_{i,j}M_{ij}c^\dagger_i c^\dagger_j|0\rangle\label{2fs}
\end{equation}
where $M$ is a $n\times n$ complex antisymmetric matrix satisfying
$\frac{1}{2}{\rm Tr}MM^\dagger=1$. As shown by Zumino in \cite{Zu.62}, for any
such matrix there is a unitary matrix  $U$ such that $U^\dagger M \bar{U}=D$,
where $D$ is a block diagonal matrix with   $2\times 2$ blocks of the form
\begin{equation}
D_k=\sqrt{\lambda_k}\begin{pmatrix}0&1\\-1&0 \end{pmatrix},
\end{equation}
with  $\lambda_k$  a real number. With the corresponding unitary transformation
$\bm{c}=U\bm{a}$ of the fermion operators, we can then split the sp space as
${\cal S}=\mathcal{S_A}\oplus\mathcal{S_B}$ and  rewrite the state (\ref{2fs})
in the normal form
\begin{equation}
|\psi\rangle=\sum_k\sqrt{\lambda_k} a^\dagger_{k(A)}a^\dagger_{k(B)}|0\rangle\label{sch}\,.
\end{equation}
Eq.\ (\ref{sch}) is the \emph{Slater decomposition} \cite{Sch.01} of
$|\psi\rangle$, a fermionic analogue (for two fermion states) of the Schmidt
decomposition for distinguishable bipartite systems. The associated one-body
density matrix is
\begin{equation}\rho^{\rm sp}=MM^\dagger=UDD^TU^\dagger\,,\end{equation}
which entails that the numbers $\lambda_k$ are its {\it eigenvalues} (twofold
degenerate), as is also evident from  Eq.\ (\ref{sch}). The Slater basis then
diagonalizes $\rho^{\rm sp}$ (and hence $\rho^{\rm qsp}$, as there are here no
pairing contractions). We then obtain $S^{\rm qsp}(|\psi\rangle)=S^{\rm
sp}(|\psi\rangle) =2\sum_k h(\lambda_k)$, which is the sum of the entanglement
entropies of all  sp  modes of the Slater basis. We also have $S_f^{\rm
sp}(|\psi\rangle)=2\sum_k f(\lambda_k)+f(1-\lambda_k)$.

Let us now consider the the one-body information loss $I_f^{\rm sp}(\rho)$ for
the ensuing mixture (\ref{W}). Since the operators $a_{k(A)}$, $a_{k(B)}$ are
the fermion operators where the one-body density matrix is diagonal, the
measurement minimizing $I_f^{\rm sp}(\rho)$ (which coincides here with $I^{\rm
qsp}(\rho)$) is that on the Slater basis of $|\psi\rangle$ and is then a
function of the eigenvalues $\lambda_k$:
\begin{eqnarray}I^{\rm sp}_f(\rho)&=&2\sum_k[ f(w\lambda_k+{\textstyle\frac{1-w}{d}})+
f(w(1-\lambda_k)+{\textstyle\frac{1-w}{d}})\nonumber\\
&&-f(w+{\textstyle\frac{1-w}{d}})-f({\textstyle\frac{1-w}{d}})]\,,\label{If2}\end{eqnarray}
with  $d=2^n$ in a grand canonical ensemble, $d=2^{n-1}$ in an even parity
ensemble and  $d=\frac{n(n-1)}{2}$ in a canonical ensemble. Present results are
then formally similar to those obtained for similar mixtures in bipartite
distinguishable systems \cite{RCC.10}, with the Slater basis replacing the
local Schmidt basis.

\section{Conclusions}

We have discussed the problem of quantifying one-body discord-like correlations
in general pure and mixed states of fermion systems, which may not have a fixed
particle number (but which commute with the number parity). First, the
correlation between a single-particle mode and its orthogonal complement in the
single-particle state space $\cal S$ is considered. The measurement of the
occupancy of a single-particle mode is properly defined and an entropic measure
of the loss of information due to this projective measurement is introduced.
The sum over all modes in a given basis of $\cal S$ of this quantity, minimized
over all such bases, is defined as the \emph{one-body information loss}, a
measure of the minimum loss of information due to an unread measurement. It is
a non-negative quantity, invariant under arbitrary unitary transformations in
$\cal S$, which vanishes if and only if the state of the system can be written
as a convex combination of Slater determinants in a common basis of $\cal S$,
i.e., if it remains invariant after the unread measurement of the occupancy of
{\it any} sp level of this basis. For pure states it reduces to the fermionic
entanglement entropy defined in \cite{GR.15}. These properties still hold if
quasiparticle level occupancy measurements are allowed, in which case
minimization is to be extended to all bases related through a Bogoliubov
transformation. The defined quantities are then extended to more general
entropic forms, including trace form entropies and  quantum Renyi entropies.
The general stationary condition to be satisfied by the minimizing measurement
was also derived.

As application, we considered the mixture of a general pure state with the
maximally mixed state for arbitrary space dimension. The minimum information
loss was shown to be always reached for a measurement on the basis
diagonalizing the generalized one-body density matrix, and to exhibit a
universal quadratic initial increase with the mixing parameter, proportional to
the quadratic entanglement entropy of the pure state. This fact implies that
it can measure non classical correlations beyond entanglement,
as explicitly verified for  ${\rm dim}({\cal S})=4$
(first non-trivial dimension for definite parity).

The approach can be applied in both canonical and grand-canonical ensembles,
and for general states such as mixtures of independent qusiparticle
states or vacua, being then suitable for studying correlations in  strongly
interacting fermion systems and requiring just a sp basis optimization.
Its  extension to more general partitions of the sp space may yield further
insight into the structure of the state and is currently under investigation.

\appendix*
\section{Entanglement of partitions of the fermionic single particle space}
We discuss here the entanglement associated to a general partition of a single
particle (sp) space ${\cal S}$ of a fermionic system. Let us assume ${\cal S}$
is of dimension $n$, and consider a partition of ${\cal S}$ in  two orthogonal
subspaces ${\cal S}_A$ and ${\cal S}_B$, generated respectively by $m$ and
$n-m$ orthogonal sp states of a given basis of ${\cal S}$,  such that  ${\cal
S}={\cal S}_A\oplus{\cal S}_B$. Any Slater determinant in this basis can then
be written, except for a global phase, as
\[|\psi_{sd}\rangle=\prod_{i\in {\cal S}_A}(c^{\dagger}_i)^{n_i}
\prod_{j\in {\cal S}_B}(c^\dagger_j)^{n_j}|0\rangle\equiv|\mu\nu\rangle\,,\]
where $n_{i(j)}=0,1$ is the occupation number of level $i$ ($j$) and
$\mu=\{n_i,i\in {\cal S}_A\},\;\nu=\{n_j,j\in{\cal S}_B\}$ indicate the
collection of these numbers. The full set of states $|\mu\nu\rangle$, with
$\mu=1,\ldots, 2^{m}$, $\nu=1\ldots,2^{n-m}$, form an orthonormal basis of the
full (grand canonical) space of many fermion states in ${\cal S}$, of dimension
$d=2^n$. Any pure fermion state $|\psi\rangle$ with definite number parity  can
then be written as
\begin{equation}
|\psi\rangle=\sum_{\mu,\nu}C_{\mu\nu}|\mu\nu\rangle\label{psimn}\,,
\end{equation}
where the sum is restricted to states $|\mu\nu\rangle$ of the same  number
parity as $|\psi\rangle$ (i.e., $|++\rangle$ or $|--\rangle$ states for
$|\psi\rangle$ of even parity). This implies that the matrix $C$ of elements
$C_{\mu\nu}$ is blocked in two submatrices ($C^+$ and $C^-$). After a singular
value decomposition $C=UDV^\dagger$,  with $D$ a ``diagonal'' $2^m\times
2^{n-m}$ matrix of non-negative elements $D_{kl}=\sigma_k\delta_{kl}$, and
$U,V$ unitary matrices, we may rewrite this state as
\begin{equation}|\psi\rangle=\sum_{k}\sigma_k |k_Ak_B\rangle,\label{psikk}\end{equation}
with $|k_Ak_B\rangle=\sum_{\mu,\nu}U_{\mu k}V^*_{\nu k}|\mu\nu\rangle$
orthonormal states. This is the Schmidt decomposition associated with this
partition. Of course, $U$ and $V$ are also blocked, so that all states
$|k_Ak_B\rangle$ have the same number parity as $|\psi\rangle$.

The entanglement entropy associated with the previous partition is then
\begin{equation}
E_{AB}=S(\rho_A)=S(\rho_B)=-\sum_{k} p_k \log_2 p_k\label{EAB},
\end{equation}
with $p_k=\sigma_k^2$ the eigenvalues of $CC^\dagger$ (or $C^\dagger C$)
and
\begin{eqnarray}\rho_A&=&\sum_{\mu,\mu'}(CC^\dagger)_{\mu\mu'}|\mu\rangle\langle\mu'|=
\sum_k p_k |k_A\rangle\langle k_B|,\\
\rho_B&=&\sum_{\nu,\nu'}(C^\dagger C)_{\nu\nu'}|\nu\rangle\langle\nu'|=\sum_k
p_k |k_B\rangle\langle k_B|,\end{eqnarray} the reduced states associated with
subspaces ${\cal S}_A$ and ${\cal S}_B$, such that any observable $O_{A(B)}$
containing operators $c^\dagger_i$, $c_i$ with $i\in {\cal S}_{A(B)}$ can be
obtained as \[\langle O_{A(B)}\rangle={\rm Tr}\rho_{A(B)} O_{A(B)}\,.\] A
Slater determinant in a given sp basis is then completely separable, in the
sense that $E_{AB}=0$ for any bipartition $(m,n-m)$ of ${\cal S}$ in this
basis. On the other hand, if a state is not a Slater determinant there is no sp
basis in which it is completely separable. Note that $\rho_{A(B)}$
commutes with number parity but will contain in general eigenstates of both
parities.

In particular, if ${\cal S}_A$ contains just a single state (say $k$), Eq.\
(\ref{psimn}) becomes
\begin{eqnarray}
|\psi\rangle&=&\sum_\nu C_{0\nu}|0\nu\rangle+\sum_{\nu'}C_{1\nu'}|1\nu'\rangle\nonumber\\
&=&\sqrt{p_{0}}\,|0\psi_0\rangle+\sqrt{p_{1}}\,|1\psi_1\rangle\label{22}
\end{eqnarray}
where $p_{i}=\sum_{\nu}|C_{i\nu}|^2$ and $|i\psi_{i}\rangle=\sum_\nu
C_{i\nu}|i\nu\rangle/\sqrt{p_{i}}$ for $i=0,1$.   Eq.\ (\ref{22}), equivalent
to (\ref{psik}), is the Schmidt decomposition of $|\psi\rangle$ since the
states $|\psi_{0}\rangle$ and $|\psi_1\rangle$ have opposite parity and are
therefore orthogonal. The ensuing entanglement (\ref{EAB})  is then given by
Eq.\ (\ref{Ek}).

Let us now consider the entanglement of mixed states. If a partition of ${\cal
S}_B$ in two subspaces ${\cal S}_{B_1}$ and ${\cal S}_{B_2}$ is made, we may
define the associated entanglement of formation of $\rho_B$, $E_{B_1
B_2}(\rho_B)$, as the minimum of the average entanglement entropies
\begin{equation}\sum_\alpha p_\alpha E_{B_1 B_2}
(|\alpha\rangle\langle\alpha|)=\sum_\alpha p_\alpha S(\rho_{B_1}^\alpha)
\end{equation}
over all decompositions $\rho_B=\sum_{\alpha}p_\alpha |\alpha\rangle\langle
\alpha|$, with $p_\alpha\geq 0$, $\sum_{\alpha}p_\alpha=1$, $\rho_{B_1}^\alpha$
the reduced state of $B_1$ in the state $|\alpha\rangle$ and  $|\alpha\rangle$
normalized many fermion states in ${\cal S}_B$, with the restriction (since
$[\rho_B,P_B]=0$) that all states $|\alpha\rangle$ have definite number parity.
This implies that $E_{B_1 B_2}$ will be the  average of the entanglement of
formations for each parity, i.e., $E_{B_1 B_2}(\rho_B)=p_+E_{B_1
B_2}(\rho_{B}^+)+p_-E_{B_1 B_2}(\rho_B^-)$ if $\rho_B=p_+\rho_B^++p_-\rho_B^-$.

In particular, if ${\cal S}_A$ is a single state, Eq.\ (\ref{22}) leads  to
\begin{equation}
\rho_B=p_0|\psi_0\rangle\langle\psi_0|+p_1|\psi_1\rangle\langle\psi_1|
\end{equation}
with $|\psi_0\rangle$ and $|\psi_1\rangle$ of opposite parity. Therefore, in
this case the decomposition is unique and the entanglement of formation reads
\begin{eqnarray}
E_{B_1 B_2}&=&p_0 S(\rho_{B_1}^0)+p_1S(\rho_{B_2}^1)=S(B_1|A)\label{Eb1}
\end{eqnarray}
where $S(B_1|A)$, equivalent to Eq.\ (\ref{Sck}),   denotes the {\it conditional} entropy
of subsystem $B_1$ after a measurement of the single level of $A$
in the mixed state $\rho_{AB_1}$, in agreement with the general result of \cite{WK.04}.

It is worth remarking that a general fermionic mixed state $\rho_{A}$ defined
over a given sp space ${\cal S}_A$ (and commuting with $P$) can be purified in
many ways, but in particular (and efficiently) by the addition of a
complementary fermionic sp space ${\cal S}_B$,  as is evident from the previous
discussion. In a grand canonical context, if $\rho_A$ is, say, of rank $2^n$,
with $2^n/2$ eigenstates of each parity, it is sufficient to add $n$ orthogonal
sp states generating an orthogonal sp space ${\cal S}_B$, and then form a pure
state like (\ref{psikk}) or (\ref{psimn}), which should have a definite number
parity if physically realizable. If all eigenstates of $\rho_A$ are of the same
parity then $n+1$ sp states should be added and just many fermion states of
definite parity should be considered.

The authors acknowledge support from CONICET (NG) and CIC (RR) of Argentina. 
Work supported by CIC and CONICET PIP 112201501-00732.

\end{document}